\documentclass{article}
\usepackage{spconf,amsmath,graphicx}
\usepackage{url}
\usepackage{booktabs}
\usepackage{multirow}
\usepackage[table]{xcolor}

\title{Multi-User Multi-Key Image Steganography with Key Isolation}
\name{Tzu-Ti Wei\sthanks{Tzu-Ti Wei and Yu-Han Tseng contributed equally to this work.}, Yu-Han Tseng\footnotemark[1], Jun-Yi Lin, Yu-Chee Tseng, Jen-Jee Chen}
\address{College of AI, National Yang Ming Chiao Tung University, Taiwan}
\begin{document}
%
\maketitle
\begin{abstract}
Steganography conceals secret information within innocuous carriers while preserving visual fidelity and enabling reliable recovery.
Recent unified networks operate normally under untriggered conditions but switch to hidden steganographic tasks when triggered.
PUSNet follows this paradigm by performing image purification during normal operation and steganographic embedding when activated.
However, it supports only a single user with one key pair, limiting its applicability in multi-user settings.
We propose \textbf{PUSNet-MK}, a multi-key extension that enforces strict key isolation via a \textbf{mismatched-key isolation loss}, effectively preventing cross-key decoding when a wrong key is applied.
This design preserves the intended steganographic behavior while addressing a critical security limitation of PUSNet.
Extensive experiments demonstrate that PUSNet-MK produces high-quality stego images and accurate secret recovery, while preventing unintended information leakage. Code is available at \url{https://gitlab.pairlab.ai.nycu.edu.tw/pairlab/ai.phd/pusnet-mk}.
\end{abstract}


\section{Introduction}
\label{sec:intro}

Steganography conceals secret information within ordinary carriers while keeping the presence of hidden content imperceptible.
Early methods rely on handcrafted embedding rules in the spatial or frequency domains~\cite{fridrich2001steganalysis,cox2007digital}, which suffer from limited capacity and weak robustness, and are vulnerable to statistical steganalysis and common image perturbations.
Recent deep learning approaches enable end-to-end optimization of hiding and recovery networks~\cite{baluja2017hiding,zhu2018hidden,wu2018steganogan,hayes2017gan,zhang2019advsteg}, learning adaptive embedding strategies that produce visually indistinguishable stego images with improved payload capacity and reliable secret extraction.
Despite these advances, most deep steganography frameworks still operate with fixed embedding or decoding configurations, limiting their flexibility in security-critical scenarios.

Among recent advances, PUSNet~\cite{li2024purified} introduces a unified framework that integrates steganographic embedding, secret recovery, and purified image reconstruction via key-activated task switching.
By using a predefined key pair, the network is triggered to perform steganographic tasks under a shared architecture.
While effective, PUSNet is restricted to a single key pair, forcing all users to share the same embedding and recovery configuration.
This design lacks user-level isolation and allows any key holder to decode secrets embedded by others, posing a critical security risk in multi-user environments.

\begin{figure}
    \centering
    \includegraphics[width=0.48\textwidth]{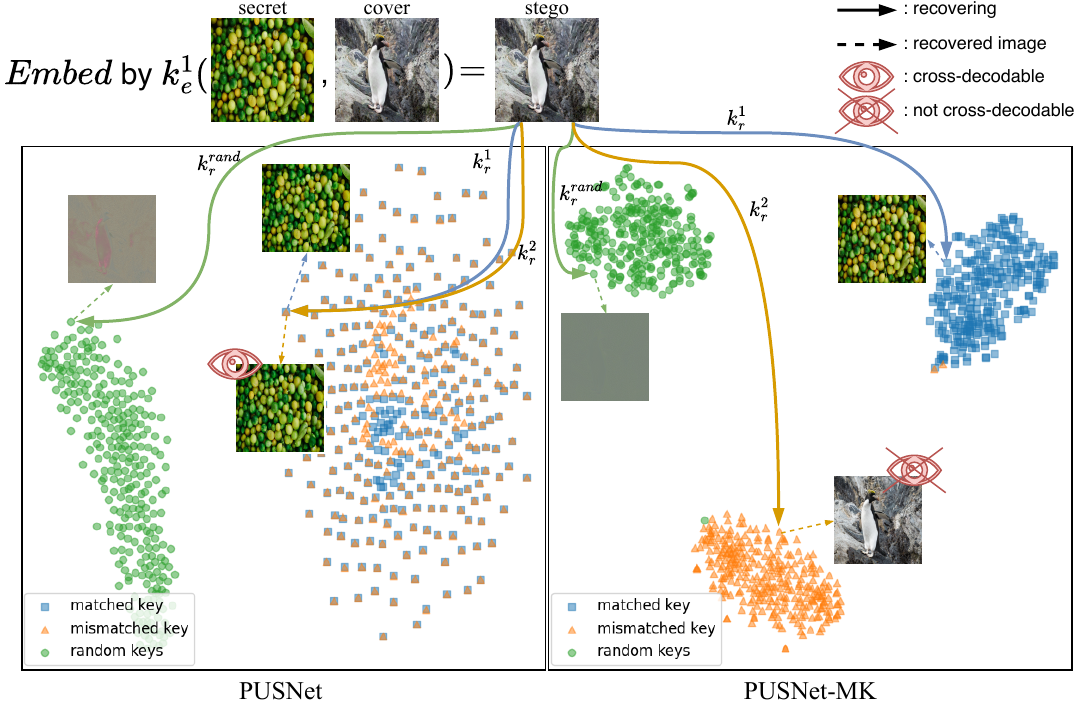}
\caption{\textbf{t-SNE.} Recovery features in PUSNet using matched and mismatched keys reveal feature entanglement. PUSNet-MK resolves this problem through key isolation.}
\label{fig:pca}
\end{figure}

In practical deployments, steganographic systems are often shared by multiple users, each requiring independent and confidential communication channels.
Formally, a multi-user steganography system must support distinct key pairs such that a secret embedded with one key can only be recovered using its corresponding key.
Failure to satisfy this property results in \emph{cross-key decoding}, where unauthorized users can partially or fully recover secrets not intended for them.
However, most existing deep steganography frameworks do not explicitly enforce key isolation.
Consequently, naively extending single-key models to multi-user settings introduces fundamental security vulnerabilities.
As shown by the t-SNE visualization in Fig.~\ref{fig:pca}, directly extending PUSNet to multiple key pairs results in entangled feature distributions during stego recovery, underscoring the need for explicit key-conditioned feature isolation to ensure secure multi-key steganography.

To address the cross-key decoding problem, we propose \textbf{PUSNet-MK}, a multi-key extension of PUSNet that enables multiple users to share a single backbone while enforcing strict key isolation.
PUSNet-MK explicitly suppresses key-induced latent interference, ensuring that only the matched key yields meaningful secret recovery.
Our contributions are summarized as follows:
\begin{itemize}
\item
We extend PUSNet to support multiple encoder--decoder key pairs within a unified backbone, enabling scalable and secure multi-user steganography.
\item
We introduce a \emph{mismatched-key isolation mechanism} that suppresses cross-key decoding by explicitly driving recoveries under mismatched keys toward non-informative (cover-like) outputs.
\item
Extensive scalability experiments demonstrate that PUSNet-MK preserves stego imperceptibility and secret fidelity comparable to the single-key baseline, while substantially strengthening key isolation.
\end{itemize}

\section{Related Works}

\subsection{Deep Learning-based Steganography}

Early image steganography methods mainly relied on hand-crafted embedding strategies in the spatial or transform domain, such as least significant bit (LSB) substitution and frequency-domain techniques~\cite{cox2007digital,fridrich2009steganography, yctseng-hiding-1, yctseng-hiding-2}.
Recent advances in deep learning have significantly improved image steganography by enabling data-driven embedding and recovery.
Baluja~\cite{baluja2017hiding} first introduced an end-to-end convolutional framework for hiding images within images.
Subsequent works further enhance imperceptibility and robustness through task-specific architectural designs, including HiDDeN~\cite{hayes2017gan}, which jointly optimizes hiding and reveal networks under adversarial constraints, and UDH~\cite{zhang2019universal}, which leverages residual learning for improved visual fidelity.
More recently, HiNet~\cite{jing2021hinet} proposes a reversible architecture to improve reconstruction accuracy and stability.
PUSNet~\cite{li2024purified} extends this line of research by integrating image purification, secret embedding, and recovery within a unified network via key-activated weight filling.
However, existing deep steganographic methods typically assume a single shared key and do not explicitly address multi-user key isolation.

\subsection{Multi-user and Multi-key Learning}
Multi-user settings have been studied in various learning systems where multiple users share a common model while requiring isolated behaviors or access control.
In such scenarios, a shared backbone is often adopted for efficiency, with user-specific conditions or parameters used to differentiate functionalities~\cite{sabour2017dynamic,perez2018film}.
However, existing designs mainly focus on improving scalability or parameter efficiency and do not explicitly address security concerns arising from user coexistence.
Recent studies have shown that shared representations can lead to unintended information leakage across conditions or users when isolation is not carefully enforced~\cite{shokri2017membership}.
In the context of steganography, this issue becomes particularly critical.
However, strict key isolation in multi-user steganographic systems remains largely unexplored.
\section{Proposed PUSNet-MK}
PUSNet~\cite{li2024purified} introduces a unified formulation for image purification, secret embedding, and secret recovery using key-conditioned sparse weight filling within a single backbone.
While effective in the single-key setting, this design does not generalize naturally to scenarios involving multiple independent keys.
In particular, when multiple keys coexist under a shared backbone, mismatched but registered keys may still partially decode hidden content, resulting in cross-decoding and unintended information leakage.
This observation indicates that the original PUSNet lacks strict key isolation and is therefore unsuitable for multi-key and multi-user deployments.
In this section, we present PUSNet-MK, a multi-key extension that preserves the efficiency of the unified backbone while enforcing robust key isolation.

\begin{figure}[!t]
    \centering
    \includegraphics[width=0.42\textwidth]{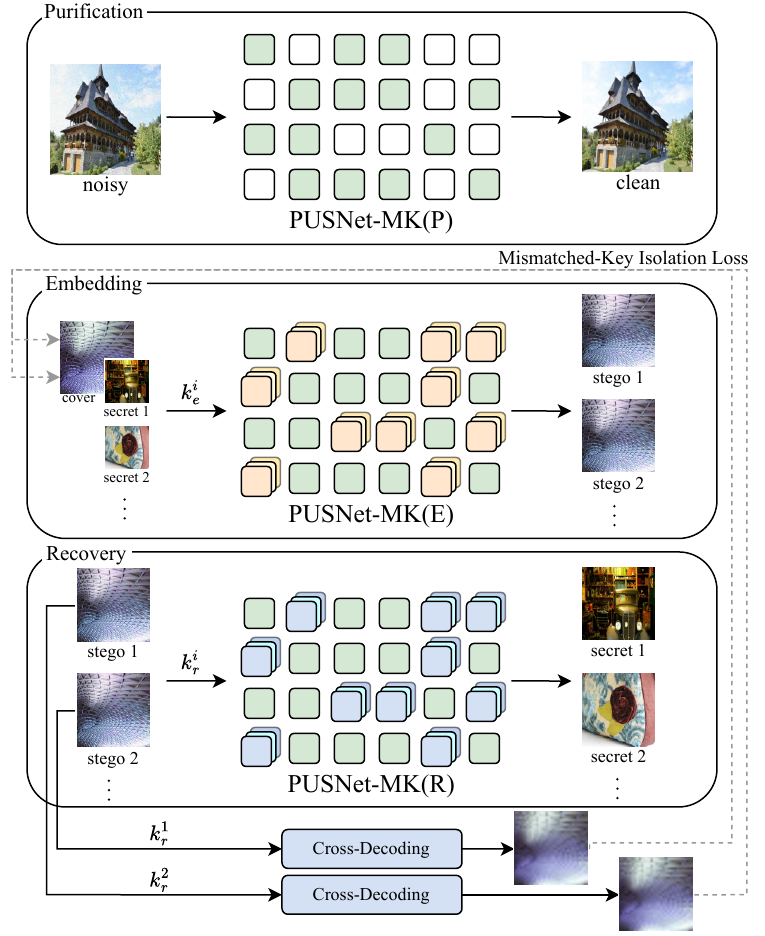}
\caption{\textbf{PUSNet-MK framework.} 
Green blocks are shared and learnable, while others are frozen and key-conditioned.}
    \label{fig:overview}
\end{figure}

\begin{table*}[!t]
    \centering
    \caption{\textbf{Quantitative comparison.} 
    We report imperceptibility ($I_{\text{cover}}$ vs. $I_{\text{stego}}$),
    recoverability ($I_{\text{secret}}$ vs. $\Tilde{I}_{\text{secret}}$), and
    cross decoding ($I_{\text{secret}}$ vs. $I^*_{\text{secret}}$), 
    where $I^*_{\text{secret}}$ means a recovered secret using a wrong key.}
    \label{tab:quantitative}
    \resizebox{\linewidth}{!}{
    \begin{tabular}{l|ccc|ccc|ccc}
        \toprule
        \multirow{2}{*}{Methods} & \multicolumn{3}{c|}{DIV2K} & \multicolumn{3}{c|}{COCO} & \multicolumn{3}{c}{ImageNet} \\
        \cline{2-10}
            & PSNR$\uparrow$ & SSIM$\uparrow$ & MAE$\downarrow$ & PSNR$\uparrow$ & SSIM$\uparrow$ & MAE$\downarrow$ & PSNR$\uparrow$ & SSIM$\uparrow$ & MAE$\downarrow$ \\ 
        \hline
            & \multicolumn{9}{c}{\textbf{imperceptibility ($I_{\text{cover}}$ vs. $I_{\text{stego}}$)}} \\
        \hline
        HiDDeN~\cite{zhu2018hidden} 
            & 24.82/21.96 & 0.817/0.742 & 0.045/0.060 
            & 25.51/22.80 & 0.831/0.758 & 0.040/0.053 
            & 25.09/22.52 & 0.823/0.746 & 0.043/0.055 \\
        Baluja~\cite{baluja2017hiding}
            & 27.06/27.41 & 0.906/0.916 & 0.036/0.035 
            & 27.86/28.30 & 0.913/0.920 & 0.032/0.031 
            & 27.13/27.63 & 0.903/0.909 & 0.036/0.035 \\
        UDH~\cite{zhang2019universal}
            & 30.18/31.35 & 0.887/0.901 & 0.024/0.021 
            & 30.56/31.93 & 0.879/0.893 & 0.023/0.020 
            & 30.35/31.68 & 0.871/0.886 & 0.024/0.020 \\
        HiNet~\cite{jing2021hinet}
            & 32.04/31.66 & 0.866/0.865 & 0.020/0.021 
            & 32.91/32.63 & 0.865/0.864 & 0.017/0.018 
            & 32.81/32.54 & 0.862/0.861 & 0.018/0.018 \\
        PUSNet~\cite{li2024purified}
            & \textbf{38.07}/\textbf{38.04} & \textbf{0.969}/\textbf{0.970} & \textbf{0.009}/\textbf{0.009}
            & \textbf{38.74}/\textbf{38.74} & \textbf{0.967}/\textbf{0.968} & \textbf{0.009}/\textbf{0.009} 
            & \textbf{38.35}/\textbf{38.37} & \textbf{0.962}/\textbf{0.962} & \textbf{0.009}/\textbf{0.009} \\
        PUSNet-MK
            & 33.77/33.65 & 0.959/0.957 & 0.017/0.017 
            & 34.47/34.40 & 0.959/0.958 & 0.015/0.015
            & 34.05/33.93 & 0.951/0.950 & 0.016/0.016 \\
        \hline
            & \multicolumn{9}{c}{\textbf{recoverability ($I_{\text{secret}}$ vs. $\Tilde{I}_{\text{secret}}$)}} \\
        \hline
        HiDDeN~\cite{zhu2018hidden} 
            & 19.68/21.92 & 0.772/0.738 & 0.081/0.063 
            & 19.89/22.52 & 0.777/0.737 & 0.080/0.056 
            & 19.88/22.15 & 0.760/0.725 & 0.081/0.059 \\
        Baluja~\cite{baluja2017hiding}
            & 28.03/27.89 & 0.909/0.904 & 0.031/0.032 
            & 28.92/28.80 & 0.913/0.908 & 0.027/0.028 
            & 28.14/28.05 & 0.902/0.898 & 0.031/0.032 \\
        UDH~\cite{zhang2019universal} 
            & 22.83/26.38 & 0.826/0.872 & 0.062/0.039 
            & 23.03/26.92 & 0.833/0.860 & 0.060/0.036 
            & 23.02/26.83 & 0.821/0.855 & 0.061/0.037 \\
        HiNet~\cite{jing2021hinet} 
            & 25.88/25.75 & 0.826/0.840 & 0.040/0.041 
            & 26.31/26.02 & 0.815/0.825 & 0.038/0.040 
            & 26.11/25.90 & 0.808/0.819 & 0.039/0.041 \\
        PUSNet~\cite{li2024purified} 
            & 29.20/29.23 & 0.947/0.948 & 0.028/0.028 
            & 29.46/29.34 & 0.947/0.947 & 0.027/0.028
            & 29.24/29.09 & 0.942/0.942 & 0.027/0.028 \\
        PUSNet-MK
            & \textbf{32.21}/\textbf{31.83} & \textbf{0.949}/\textbf{0.948} & \textbf{0.019}/\textbf{0.020} 
            & \textbf{33.17}/\textbf{32.79} & \textbf{0.952}/\textbf{0.951} & \textbf{0.017}/\textbf{0.018} 
            & \textbf{32.64}/\textbf{32.28} & \textbf{0.946}/\textbf{0.943} & \textbf{0.018}/\textbf{0.019} \\
        \hline
            & \multicolumn{9}{c}{\textbf{cross decoding ($I_{\text{secret}}$ vs. $I^*_{\text{secret}}$)}} \\
        \hline
        HiDDeN~\cite{zhu2018hidden} 
            & 19.21/20.67 & 0.740/0.730 & 0.092/0.071 
            & 19.07/21.18 & 0.740/0.730 & 0.094/0.064 
            & 19.03/20.89 & 0.720/0.720 & 0.094/0.067 \\
        Baluja~\cite{baluja2017hiding}
            & 19.21/20.00 & 0.771/0.792 & 0.086/0.080 
            & 19.06/19.70 & 0.774/0.788 & 0.086/0.082 
            & 18.88/19.43 & 0.764/0.779 & 0.089/0.085 \\
        UDH~\cite{zhang2019universal}
            & 25.46/22.30 & 0.880/0.830 & 0.043/0.065
            & 25.53/22.36 & 0.870/0.830 & 0.043/0.065
            & 25.70/22.42 & 0.870/0.820 & 0.042/0.065 \\
        HiNet~\cite{jing2021hinet}
            & 25.88/18.00 & 0.830/0.430 & 0.040/0.102 
            & 26.31/18.06 & 0.820/0.380 & 0.038/0.102 
            & 26.11/18.00 & 0.810/0.380 & 0.039/0.102 \\
        PUSNet~\cite{li2024purified}
            & 29.53/28.36 & 0.945/0.943 & 0.026/0.031
            & 29.81/28.55 & 0.945/0.942 & 0.026/0.030 
            & 29.43/28.46 & 0.940/0.938 & 0.027/0.030 \\
        PUSNet-MK
            &  \textbf{8.67}/\textbf{8.77}  & \textbf{0.153}/\textbf{0.161} & \textbf{0.312}/\textbf{0.303}
            &  \textbf{8.51}/\textbf{8.52}  & \textbf{0.215}/\textbf{0.214} & \textbf{0.315}/\textbf{0.315}
            &  \textbf{8.48}/\textbf{8.50}  & \textbf{0.207}/\textbf{0.208} & \textbf{0.318}/\textbf{0.318} \\
            \bottomrule
    \end{tabular}
    }
\end{table*}

\subsection{Review of PUSNet}

Fig.~\ref{fig:overview} illustrates the PUSNet-MK framework.
When only a single key pair is used, it is equivalent to PUSNet.
PUSNet supports image purification (\textbf{P}), secret embedding (\textbf{E}), and secret recovery (\textbf{R}) by activating different subsets of network parameters via a sparse weight filling mechanism.

PUSNet first initializes a backbone weight set $W$ and samples a random binary mask $M$ according to a predefined sparse ratio $\alpha$.
The masked weights $W \odot M$ form a shared backbone across all tasks, while the complementary region $W \odot \overline{M}$ is reserved for task-specific adaptations.
This design enables multiple functionalities to be realized within a single network without additional task-specific modules.

For image purification, the purified network operates as
\begin{equation}
    N[W \odot M + W \odot \overline{M}](I_{\text{noisy}}) \rightarrow \tilde{I}_{\text{clean}},
\end{equation}
where $I_{noisy}$ is derived from a clean image $I_{clean}$ by adding Gaussian noise.
During training, the optimization is restricted to the shared subset $W \odot M$ and the complementary subset $W \odot \overline{M}$ is frozen and excluded from parameter updates.

For steganographic embedding and recovery, PUSNet employs key-conditioned weight instantiation.
Given a pair of keys $(k_e, k_r)$, a weight set $W_e$ is generated by seed $k_e$ and another $W_r$ is generated by seed $k_r$ via weights initialization method~\cite{glorot2010understanding}.
Weights $W_e$ and $W_r$ populate the complementary region $\overline{M}$ to perform secret embedding and recovering tasks:
\begin{equation}
    \begin{aligned}
        N[W \odot M + W_e \odot \overline{M}](I_{\text{secret}}, I_{\text{cover}}) &\rightarrow I_{\text{stego}} \\
        N[W \odot M + W_r \odot \overline{M}](I_{\text{stego}}) &\rightarrow \Tilde{I}_{\text{secret}}
    \end{aligned}
    \label{eq:encode-decode}
\end{equation}
where $I_{\text{secret}}$ and $I_{\text{cover}}$ denote the secret and cover images, respectively.
$I_{\text{stego}}$ is the stego image obtained by embedding $I_{\text{secret}}$ into $I_{\text{cover}}$, which is expected to be visually indistinguishable from $I_{\text{cover}}$.
After recovery, the reconstructed secret $\Tilde{I}_{\text{secret}}$ should be visually similar to $I_{\text{secret}}$.
Again, during training, only the shared subset $W \odot M$ is optimized.

\subsection{Multi-key Extension and Training Objectives}

To support multiple key pairs, we preserve the original PUSNet architecture and extend it to the multi-key setting by enforcing strong key-conditioned separation in the recovery feature space as shown in Fig.~\ref{fig:overview}.
The subnetwork $W \odot M$ remains fully shared and jointly optimized across all normal and steganographic tasks, while each key pair activates its own embedding and recovery functions via its corresponding complementary subnetworks.
This design enables multiple users to coexist within a unified backbone.

In the multi-key setting, PUSNet-MK allows $K$ key pairs $\{(k_e^i, k_r^i)\}_{i=1}^{K}$.
For each pair $(k_e^i, k_r^i)$, similar weight sets $W_e^i$ and $W_r^i$ are generated using keys as seeds.
The secret embedding and recovering tasks are performed by:
\begin{equation}
\begin{aligned}
N[W \odot M + W_e^i \odot \overline{M}]
(I^i_{\text{secret}}, I^i_{\text{cover}}) &\rightarrow I^i_{\text{stego}} 
\\
N[W \odot M + W_r^i \odot \overline{M}]
(I^i_{\text{stego}}) &\rightarrow \Tilde{I}^i_{\text{secret}}
\end{aligned}
\label{eq:encode-decode-mk}
\end{equation}
The training remains the same---only the shared subset $W \odot M$ is optimized.

\begin{figure*}[!t]
    \centering
    \includegraphics[width=0.90\textwidth]{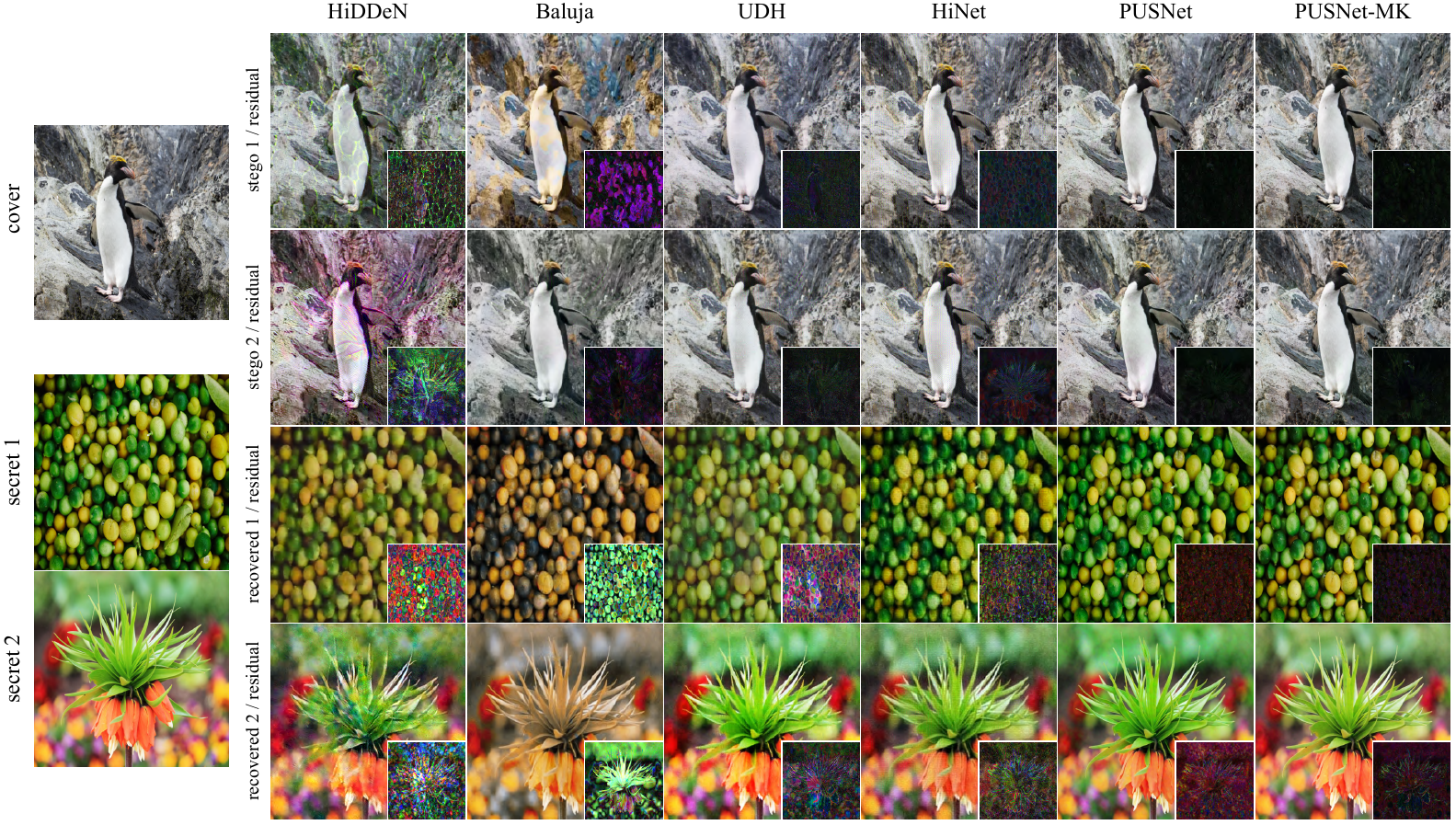}
\caption{\textbf{Qualitative comparison.}
Stego and recovered images, with residuals shown in the lower-right corner.}
    \label{fig:qualitative}
\end{figure*}

\textbf{Cross-Key Decoding Problem.}
As shown in Eq.~\ref{eq:encode-decode-mk}, all key pairs operate on the same shared backbone (i.e., $M$), while realizing distinct functionalities through their complementary regions (i.e., $\overline{M}$).
Although this design enables unification within a single network, it also introduces the risk of \emph{key interference}, where a mismatched key can partially recover recognizable secret content, thereby violating the security requirement of multi-user steganography.
As illustrated in Fig.~\ref{fig:pca}, we embed a secret using one key and attempt recovery with multiple keys.
Without explicit key separation, the recovery features corresponding to matched and mismatched keys exhibit significant overlap, leading to unintended cross-key decoding and secret leakage. To meet the security requirements of multi-user steganography, we next present a solution that enforces strict separation among key-conditioned representations.

\subsection{Training Objective and Key Separation}

Following PUSNet, we jointly optimize the normal, steganographic embedding, and secret recovery tasks across all key pairs.
In addition, we introduce a mismatched-key isolation loss to explicitly enforce strict key separation.
The overall training objective promotes imperceptible stego generation, faithful secret recovery under matched keys, preserved purification capability, and non-informative outputs when mismatched keys are used.

\textbf{Steganographic Losses.}
For each key pair $(k_e^i, k_r^i)$, $i=1..K$, the secret embedding and recovery results are optimized by:
\begin{equation}
\begin{aligned}
    \mathcal{L}_{\text{emb}} &=
    \sum_{i=1..K} \| I^i_{\text{stego}} - I^i_{\text{cover}} \|^2
    \\
    \mathcal{L}_{\text{rec}} &=
    \sum_{i=1..K} \| \Tilde{I}^i_{\text{secret}} - I^i_{\text{secret}} \|^2 
    .
\end{aligned}
\end{equation}

\textbf{Purification Loss.}
To preserve the purification capability, the normal network is optimized by :
\begin{equation}
\mathcal{L}_{\text{pur}} =
\| I_{\text{clean}} - \tilde{I}_{\text{clean}} \|^2 .
\end{equation}

\textbf{Mismatched-Key Isolation Loss.}
To suppress cross-key decoding, a stego imaged produced by a key pair $(k_e^i, k_r^i)$ but recovered by a mismatched key $(k_e^j, k_r^j)$, $i \neq j$, is forced to regress toward the cover image $I_{\text{cover}}$:
\begin{equation}
\mathcal{L}_{\text{mki}} =
\sum_{i \neq j} 
\| N[W \odot M + W_r^j \odot \overline{M}]
(I_{\text{stego}}^i) - I_{\text{cover}} \|^2 .
\end{equation}

\textbf{Total Loss.}
The final objective is to optimize:
\begin{equation}
\mathcal{L}_{\text{total}} =
\lambda_e \mathcal{L}_{\text{emb}} +
\lambda_r \mathcal{L}_{\text{rec}} +
\lambda_p \mathcal{L}_{\text{pur}} +
\lambda_m \mathcal{L}_{\text{mki}}.
\label{eq:total-loss}
\end{equation}
$\mathcal{L}_{\text{total}}$ is used to train the shared weights in $M$.
For each component in Eq.~\ref{eq:total-loss}, the corresponding complementary weights in $\overline{M}$ should be instantiated accordingly.
An ablation study on the effect of the mismatched-key isolation weight $\lambda_m$ is provided in the supplementary material.
\section{Experiments}
All models are trained on DIV2K~\cite{Agustsson2017DIV2K} with images resized to $256 \times 256$ for 500 epochs using a single NVIDIA L40S GPU.
We evaluate sparse ratios in the range $\alpha \in [0.5, 0.9]$.
The original PUSNet with a direct multi-key extension serves as the primary baseline.
Evaluation is conducted on DIV2K, COCO~\cite{lin2014coco}, and ImageNet~\cite{deng2009imagenet}, with all images resized to $512 \times 512$.
The hyperparameters are set to $\lambda_e = 1.0$, $\lambda_r = 0.75$, $\lambda_p = 0.25$, and $\lambda_m = 0.5$.

\textbf{Quantitative Comparison.}
Table~\ref{tab:quantitative} reports quantitative results on \emph{imperceptibility} (hiding quality), \emph{recoverability} (secret fidelity), and \emph{cross-key decoding} (leakage under mismatched keys), evaluated using PSNR, SSIM, and MAE.
PUSNet-MK outperforms all competing methods in terms of recoverability and achieves substantially lower leakage in cross-key decoding with large margins.
While the original PUSNet attains the best imperceptibility due to the absence of constraints on the recovery feature space, PUSNet-MK remains highly competitive, particularly in SSIM and MAE.
These results confirm that PUSNet-MK effectively supports secure multi-key image steganography.

\textbf{Qualitative Comparison.}
Fig.~\ref{fig:qualitative} presents qualitative comparisons under matched key settings.
Most recent methods generate visually imperceptible stego images.
As evidenced by the residual images, PUSNet-MK reconstructs secrets with the highest visual fidelity while maintaining competitive imperceptibility.

\begin{figure}
    \centering
    \includegraphics[width=0.48\textwidth]{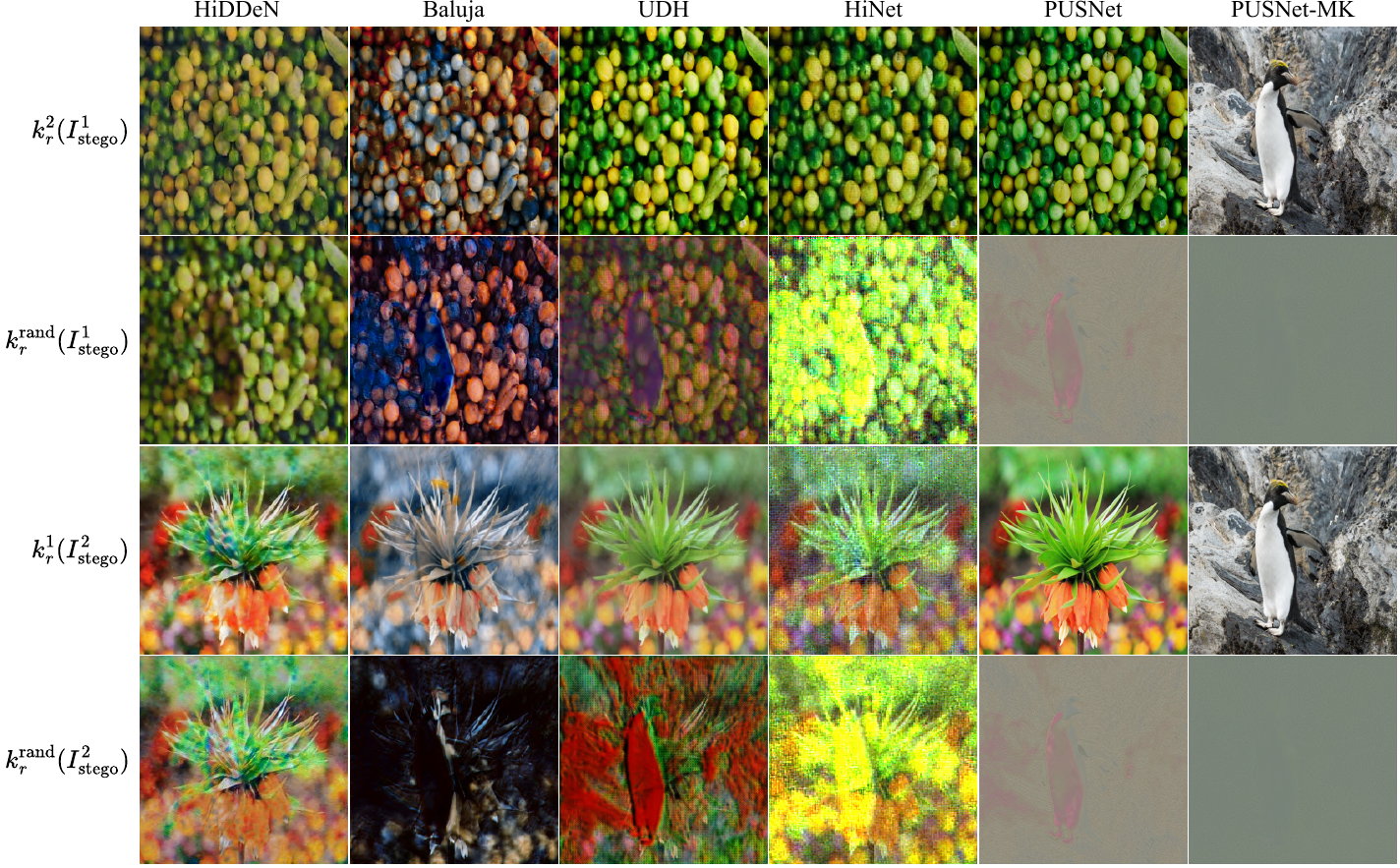}
\caption{\textbf{Cross-key decoding tests.} 
$k^j_r (I_{\text{stego}}^{i}) = N[W \odot M + W_r^{j} \odot \overline{M}](I_{\text{stego}}^{i})$ denotes decoding $I_{\text{stego}}^{i}$ with a mismatched key $k^j_r$, $i \neq j$, where $rand$ means a random key.}
    \label{fig:cross-key}
\end{figure}

\textbf{Cross-Key Decoding Test.}
Using the scenarios in Fig.~\ref{fig:qualitative}, Fig.~\ref{fig:cross-key} presents cross-key decoding results.
Most baselines recover noticeable secret structures under either mismatched or random keys, indicating insufficient key isolation.
PUSNet reveals noticeable RGB information when a mismatched registered key is used and leaks minor structural information under a random key.
In contrast, PUSNet-MK produces non-informative outputs under random keys and cover-like outputs under mismatched keys.
These results demonstrate that PUSNet-MK effectively enforces strict key-conditioned separation and prevents unintended secret leakage.

\begin{figure}
    \centering
    \includegraphics[width=0.4\textwidth]{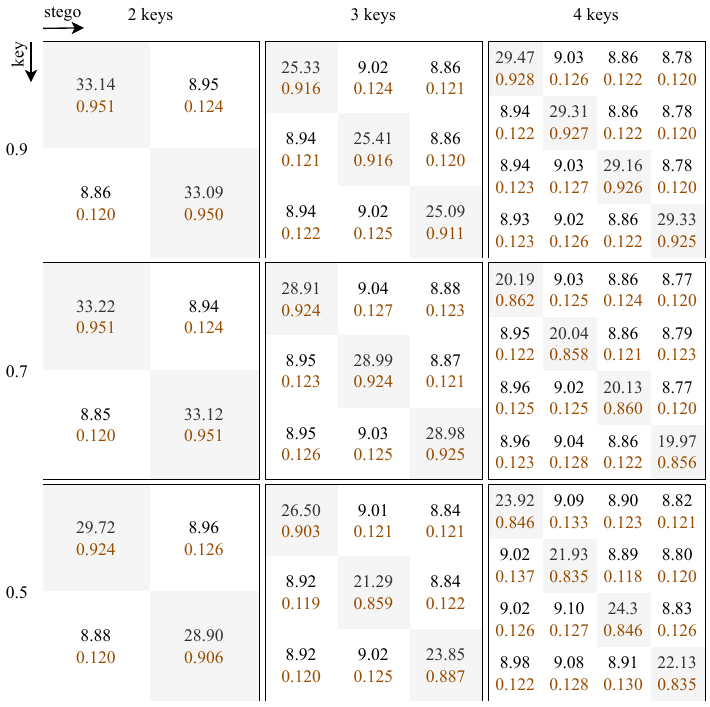}
\caption{\textbf{Scalability study.}
We vary the number of keys ($K$) and sparse ratio ($\alpha$).
Each confusion matrix shows recoverability, where rows correspond to stego images encoded with different keys and columns correspond to keys used for decoding.
In each cell, we report PSNR (black) and SSIM (brown).}
    \label{fig:multi-key}
\end{figure}

\textbf{Scalability ($K$, $\alpha$).}
Fig.~\ref{fig:multi-key} investigate the scalability of PUSNet-MK in terms of recoverability by varying $K$ and ratio $\alpha$.
Across different $\alpha$, matched-key decoding consistently achieves high reconstruction quality, as reflected by the strong diagonal responses.
In contrast, cross-key decoding remains uniformly suppressed, with off-diagonal entries staying at a low level regardless of the number of keys.
These results demonstrate its robustness and scalability for multi-user steganography.

\section{Conclusions}
We identify cross-key decoding as a critical limitation of single-key steganographic frameworks in multi-user settings.
To address this issue, we propose PUSNet-MK, a multi-key extension that allows multiple users to share a single backbone while enforcing strict key isolation.
Experiments on three public datasets demonstrate that PUSNet-MK preserves steganographic imperceptibility and matched-key recovery performance, while effectively suppressing cross-key decoding.
These results validate the practicality of PUSNet-MK for secure and scalable multi-user steganography. 
A limitation of the current framework is its vulnerability to steganalysis. Although PUSNet-MK improves key isolation, detection by SRNet~\cite{boroumand2018deep} remains feasible: the detection probability stays below 60\% when fewer than 30 stego images are exposed, but exceeds 70\% when more than 40 stego images are available, indicating limited robustness under larger-scale sample exposure.



\bibliographystyle{IEEEbib}
\bibliography{refs}

@String(CVPR= {IEEE Conf. Comput. Vis. Pattern Recog.})

@String(ICCV= {Int. Conf. Comput. Vis.})

@String(ECCV= {Eur. Conf. Comput. Vis.})

@String(AAAI = {AAAI})

@String(CVPRW= {IEEE Conf. Comput. Vis. Pattern Recog. Worksh.})

@String(CVPR  = {CVPR})

@String(ICCV  = {ICCV})

@String(ECCV  = {ECCV})

@String(CVPRW= {CVPRW})

@ARTICLE{yctseng-hiding-1,
  author={Yu-Chee Tseng and Yu-Yuan Chen and Hsiang-Kuang Pan},
  journal={IEEE Transactions on Communications}, 
  title={A secure data hiding scheme for binary images}, 
  year={2002},
  volume={50},
  number={8},
  pages={1227-1231}
}

@ARTICLE{yctseng-hiding-2,
  author={Yu-Chee Tseng and Hsiang-Kuang Pan},
  journal={IEEE Transactions on Computers}, 
  title={Data hiding in 2-color images}, 
  year={2002},
  volume={51},
  number={7},
  pages={873-880}
}

@article{hayes2017gan,
  title   = {Generating Steganographic Images via Adversarial Training},
  author  = {Hayes, Jamie and Danezis, George},
  journal = {arXiv preprint arXiv:1703.00371},
  year    = {2017}
}

@article{zhang2019advsteg,
  title   = {AdvSteg: Adversarial Steganography via Generative Adversarial Networks},
  author  = {Zhang, Ruoyu and Dong, Shunquan and Liu, Jun and Wang, Yi},
  journal = {IEEE Access},
  volume  = {7},
  pages   = {110038--110047},
  year    = {2019}
}

@article{zhu2018hidden,
  title   = {HiDDeN: Hiding Data With Deep Networks},
  author  = {Zhu, Jiren and Kaplan, Russell and Johnson, Justin and Li, Fei-Fei},
  journal = {arXiv preprint arXiv:1807.09937},
  year    = {2018}
}

@inproceedings{baluja2017hiding,
  title     = {Hiding images in plain sight: Deep steganography},
  author    = {Baluja, Shumeet},
  booktitle = {Advances in Neural Information Processing Systems (NeurIPS)},
  pages     = {2068--2077},
  year      = {2017}
}

@inproceedings{zhang2019universal,
  title     = {Universal deep hiding for steganography},
  author    = {Zhang, Rui and Dong, Xia and Zhang, Xinpeng},
  booktitle = {Proceedings of the ACM International Conference on Multimedia (ACM MM)},
  pages     = {2567--2575},
  year      = {2019}
}

@inproceedings{jing2021hinet,
  title     = {HiNet: Deep image hiding by invertible network},
  author    = {Jing, Yue and Yang, Yujie and Liu, Zongwei and Tian, Yong and Sebe, Nicu},
  booktitle = {Proceedings of the IEEE/CVF International Conference on Computer Vision (ICCV)},
  pages     = {4733--4742},
  year      = {2021}
}

@inproceedings{li2024purified,
  title     = {Purified and Unified Steganographic Network},
  author    = {Li, Guobiao and Li, Sheng and Luo, Zicong and Qian, Zhenxing and Zhang, Xinpeng},
  booktitle = {Proceedings of the IEEE/CVF Conference on Computer Vision and Pattern Recognition (CVPR)},
  pages     = {27569--27578},
  year      = {2024}
}

@article{fridrich2001steganalysis,
  title={Reliable detection of LSB steganography in color and grayscale images},
  author={Fridrich, Jessica and Goljan, Miroslav and Du, Rui},
  journal={Proceedings of the 2001 Workshop on Multimedia and Security},
  pages={27--30},
  year={2001},
  organization={ACM}
}

@book{cox2007digital,
  title={Digital Watermarking and Steganography},
  author={Cox, Ingemar J. and Miller, Matthew L. and Bloom, Jeffrey A. and Fridrich, Jessica and Kalker, Ton},
  publisher={Morgan Kaufmann},
  edition={2nd},
  year={2007}
}

@inproceedings{wu2018steganogan,
  title={SteganoGAN: High capacity image steganography with GANs},
  author={Wu, Shumei and Shi, Haichao and Dong, Jian and Su, Zhihua and Xue, Hui},
  booktitle={Proceedings of the IEEE/CVF Conference on Computer Vision and Pattern Recognition Workshops (CVPRW)},
  pages={1682--1685},
  year={2018},
  organization={IEEE}
}

@inproceedings{lin2014coco,
  title     = {Microsoft COCO: Common Objects in Context},
  author    = {Lin, Tsung-Yi and Maire, Michael and Belongie, Serge and Hays, James
               and Perona, Pietro and Ramanan, Deva and Doll{\'a}r, Piotr and Zitnick, C. Lawrence},
  booktitle = {European Conference on Computer Vision (ECCV)},
  pages     = {740--755},
  year      = {2014},
  publisher = {Springer}
}

@inproceedings{Agustsson2017DIV2K,
  title     = {NTIRE 2017 Challenge on Single Image Super-Resolution: Dataset and Study},
  author    = {Agustsson, Eirikur and Timofte, Radu},
  booktitle = {IEEE Conference on Computer Vision and Pattern Recognition Workshops (CVPRW)},
  year      = {2017},
  note      = {Introduces the DIV2K dataset}
}

@inproceedings{deng2009imagenet,
  title     = {ImageNet: A Large-Scale Hierarchical Image Database},
  author    = {Deng, Jia and Dong, Wei and Socher, Richard and Li, Li-Jia
               and Li, Kai and Fei-Fei, Li},
  booktitle = {IEEE Conference on Computer Vision and Pattern Recognition (CVPR)},
  pages     = {248--255},
  year      = {2009}
}

@inproceedings{glorot2010understanding,
  title={Understanding the difficulty of training deep feedforward neural networks},
  author={Glorot, Xavier and Bengio, Yoshua},
  booktitle={Proceedings of the 13th International Conference on Artificial Intelligence and Statistics (AISTATS)},
  pages={249--256},
  year={2010}
}

@book{fridrich2009steganography,
  title={Steganography in digital media: principles, algorithms, and applications},
  author={Fridrich, Jessica},
  year={2009},
  publisher={Cambridge university press}
}

@article{sabour2017dynamic,
  title={Dynamic routing between capsules},
  author={Sabour, Sara and Frosst, Nicholas and Hinton, Geoffrey E},
  journal={Advances in neural information processing systems},
  volume={30},
  year={2017}
}

@inproceedings{perez2018film,
  title={Film: Visual reasoning with a general conditioning layer},
  author={Perez, Ethan and Strub, Florian and De Vries, Harm and Dumoulin, Vincent and Courville, Aaron},
  booktitle={Proceedings of the AAAI conference on artificial intelligence},
  volume={32},
  number={1},
  year={2018}
}

@inproceedings{shokri2017membership,
  title={Membership inference attacks against machine learning models},
  author={Shokri, Reza and Stronati, Marco and Song, Congzheng and Shmatikov, Vitaly},
  booktitle={2017 IEEE symposium on security and privacy (SP)},
  pages={3--18},
  year={2017},
  organization={IEEE}
}

@article{boroumand2018deep,
  title={Deep residual network for steganalysis of digital images},
  author={Boroumand, Mehdi and Chen, Mo and Fridrich, Jessica},
  journal={IEEE Transactions on Information Forensics and Security},
  volume={14},
  number={5},
  pages={1181--1193},
  year={2018},
  publisher={IEEE}
}
\newpage
\onecolumn
\setcounter{section}{0}
\setcounter{figure}{0}
\setcounter{table}{0}
\setcounter{equation}{0}
\renewcommand{\thefigure}{S\arabic{figure}}
\renewcommand{\thetable}{S\arabic{table}}
\renewcommand{\theequation}{S\arabic{equation}}

\section{Supplementary Material}

\subsection{Ablation on the mismatched-key isolation weight}
Table~\ref{tab:lambda-m-ablation} studies the effect of the mismatched-key isolation weight $\lambda_m$.
When $\lambda_m=0$, the model achieves reasonable imperceptibility and recoverability, but cross-key decoding remains severe, indicating that the recovered secret still contains substantial leaked information under mismatched keys.
Once the isolation loss is enabled, the cross-decoding performance drops sharply, showing that even a moderate $\lambda_m$ is sufficient to suppress key interference.

Among the tested settings, $\lambda_m=0.5$ provides the best overall trade-off.
It achieves the strongest imperceptibility and recoverability in terms of PSNR and MAE, while already reducing cross-key decoding to the same low level as larger $\lambda_m$ values.
Although increasing $\lambda_m$ to $0.7$ or $1.0$ yields slightly stronger suppression in some cross-decoding metrics, it also degrades matched-key recovery and stego quality.
Based on this trade-off, we set $\lambda_m=0.5$ in the main experiments.

\begin{table}[h]
    \centering
    \caption{\textbf{Ablation on the mismatched-key isolation weight $\lambda_m$.}
    We evaluate different choices of $\lambda_m$ on DIV2K in terms of imperceptibility ($I_{\text{cover}}$ vs. $I_{\text{stego}}$), recoverability ($I_{\text{secret}}$ vs. $\Tilde{I}_{\text{secret}}$), and cross decoding ($I_{\text{secret}}$ vs. $I^*_{\text{secret}}$), where $I^*_{\text{secret}}$ denotes the secret recovered with a mismatched key.}
    \label{tab:lambda-m-ablation}
    \begin{tabular}{l|ccc}
        \toprule
        \multirow{2}{*}{Methods} & \multicolumn{3}{c}{DIV2K} \\
        \cline{2-4}
            & PSNR$\uparrow$ & SSIM$\uparrow$ & MAE$\downarrow$ \\ 
        \hline
            & \multicolumn{3}{c}{\textbf{imperceptibility ($I_{\text{cover}}$ vs. $I_{\text{stego}}$)}} \\
        \hline
        $\lambda_m=0.0$
            & 30.74/30.87 & 0.951/0.948 & 0.022/0.022 \\
        $\lambda_m=0.3$
            & 31.86/31.75 & 0.950/0.949 & 0.019/0.019 \\
        $\lambda_m=0.5$
            & \textbf{33.47}/\textbf{33.79} & 0.941/0.941 & \textbf{0.015}/\textbf{0.015} \\
        $\lambda_m=0.7$
            & 31.62/31.67 & \textbf{0.955}/\textbf{0.955} & 0.019/0.019 \\
        $\lambda_m=1.0$
            & 29.45/29.44 & 0.940/0.938 & 0.025/0.025 \\
        \hline
            & \multicolumn{3}{c}{\textbf{recoverability ($I_{\text{secret}}$ vs. $\Tilde{I}_{\text{secret}}$)}} \\
        \hline
        $\lambda_m=0.0$
            & 30.10/30.30 & 0.938/0.935 & 0.024/0.023 \\
        $\lambda_m=0.3$
            & 31.19/30.91 & \textbf{0.942}/\textbf{0.940} & 0.021/0.021 \\
        $\lambda_m=0.5$
            & \textbf{31.68}/\textbf{31.48} & 0.937/0.936 & \textbf{0.019}/\textbf{0.020} \\
        $\lambda_m=0.7$
            & 29.39/29.26 & 0.934/0.926 & 0.026/0.027 \\
        $\lambda_m=1.0$
            & 25.77/25.68 & 0.903/0.898 & 0.040/0.040 \\
        \hline
            & \multicolumn{3}{c}{\textbf{cross decoding ($I_{\text{secret}}$ vs. $I^*_{\text{secret}}$)}} \\
        \hline
        $\lambda_m=0.0$
            & 30.15/29.14 & 0.933/0.925 & 0.024/0.026 \\
        $\lambda_m=0.3$
            &  8.55/ 8.79 & 0.149/0.146 & 0.312/0.295 \\
        $\lambda_m=0.5$
            &  8.54/ 8.79 & 0.149/0.147 & {0.312}/0.295 \\
        $\lambda_m=0.7$
            &  8.54/ 8.80 & 0.150/0.148 & 0.312/0.295 \\
        $\lambda_m=1.0$
            &  \textbf{8.52}/ \textbf{8.75} & \textbf{0.148}/\textbf{0.146} & \textbf{0.313}/\textbf{0.296}  \\
            \bottomrule
    \end{tabular}
\end{table}
\end{document}